\documentclass[12pt]{article}
\newcommand{\Frac}[2]{\frac{\displaystyle #1}{\displaystyle #2}}
\newcommand{\beq}{\begin{equation}}
\newcommand{\eeq}{\end{equation}}
\newcommand{\beqn}{\begin{eqnarray}}
\newcommand{\eeqn}{\end{eqnarray}}
\newcommand{\beqns}{\begin{eqnarray*}}
\newcommand{\eeqns}{\end{eqnarray*}}

\begin{document}
\begin{titlepage}
\begin{center}
\hfill  hep-ph/0202002
 \vskip2.5cm
 {\Large {\bf The CP-violating asymmetry in $\eta\rightarrow\pi^+
\pi^- e^+e^-$ \\}} \vspace*{1.5cm}
 Dao-Neng Gao$^\dagger$\\
\vspace*{0.4cm} {\it Department of Astronomy and Applied Physics,
University of Science and Technology of China, Hefei, Anhui
230026, China}
 \vspace*{1cm}
\end{center}
\begin{abstract}
\noindent 
We study the CP-violating asymmetry ${\cal A}_{\rm CP}$, which
arises, in $\eta\rightarrow\pi^+\pi^- e^+ e^-$, from the angular
correlation of the $e^+ e^-$ and $\pi^+\pi^-$ planes due to the
interference between the magnetic and electric decay amplitudes. With the
phenomenologically determined magnetic amplitude and branching
ratio as input, the asymmetry, induced by the electric
bremsstrahlung amplitude through the CP-violating decay
$\eta\rightarrow\pi^+\pi^-$, and by an unconventional tensor type
operator, has been estimated respectively. The upper bound of 
${\cal A}_{\rm CP}$ from the former is about $10^{-3}$, and the 
asymmetry from the latter might be up to $O(10^{-2})$. One can
therefore expect that this CP asymmetry would be an interesting
CP-violating observable for the future precise measurements in the $\eta$
factories.
\end{abstract}

\vfill \noindent
$^\dagger$ E-mail:~gaodn@ustc.edu.cn
\end{titlepage}

 CP violation has been observed experimentally in the
flavor-changing weak decays of the neutral $K$-mesons
\cite{CCFT64,NA48,KTeV} and $B$-mesons \cite{Babar,Belle}.
However, the origin of the violation remains unclear in the modern
particle physics. The standard model predicts that the only way
that CP is violated is through the Cabbibo-Kobayashi-Maskawa (CKM)
mechanism \cite{CKM}. Specifically, the source of CP violation is
a single phase in the CKM mixing matrix that describes the
flavor-changing weak interaction couplings of quarks. Although the
predictions based on CKM mechanism are consistent with the
observations in $K$ and $B$ systems, it would be interesting to
look for the other sources of CP violation beyond the CKM phase
and outside flavor-changing processes, in order both to increase
our understanding in the CP violation itself and to look for new
physics effects beyond the standard model \cite{Nir}.

Rare $\eta$ decays provide a good laboratory for the above motivations,
and CP violation in some rare $\eta$ decays has been explored by
experimentalists \cite{Kull98}. Theorectally, 
P violation and CP violation in the decay of $\eta\to\pi^+\pi^-\gamma$
have been discussed by Herczeg and Singer \cite{HP73} nearly thirty years
before; very recently, Geng, Ng, and Wu \cite{GNW02} studied the
CP-violating effects in this decay by considering the photon
polarizations, and they predicted that a sizable linear photon
polarization could be expected in some unconventional new physics
scenarios. In the neutral $K$ system, a large CP asymmetry, which arises
from the interference between the parity-conserving magnetic amplitudes
and the parity-violating electric amplitudes of $K_L\rightarrow
\pi^+\pi^-\gamma^*\rightarrow\pi^+\pi^- e^+ e^-$, has been
predicted theoretically by many authors \cite{SW,ESW,EP01} and
confirmed experimentally \cite{KTeV00, NA48}. The purpose of this
paper is to extend the above analyses to the decay of
$\eta\rightarrow\pi^+\pi^-\gamma^*\rightarrow\pi^+\pi^-e^+ e^-$, thus to
probe possible new CP-violating effects in $\eta$ factories.
It will be shown below that an interesting CP-violating observable
could be constructed if a relevant parity-violating electric
transition exists.

The invariant decay amplitude of $\eta(p)\rightarrow
\pi^+(p_+)\pi^-(p_-) e^+(k_+) e^-(k_-)$ can be expressed as
follows
\beqn\label{Amp1}
A(\eta\rightarrow\pi^+\pi^- e^+
e^-)=\frac{e}{q^2}\bar{u}(k_-)\gamma_\mu v(k_+)
\left(M\varepsilon^{\mu\nu\alpha\beta} p_{+\nu} p_{-\alpha}
q_{\beta}+E_+ p_+^\mu+ E_- p_-^\mu
 \right),
\eeqn where $q=k_++k_-$. The Lorentz invariant form factors $M$
and $E_\pm$ stand for parity-conserving magnetic and
parity-violating electric transitions respectively, which depend
on scalar products of $q$, $p_+$ and $p_-$.  In order to discuss
CP violation in $\eta\rightarrow\pi^+\pi^- e^+ e^-$, it is
convenient to use the same kinematic variables as those in
$K_L\rightarrow\pi^+\pi^- e^+ e^-$, which have been used by Pais
and Trieman for semileptonic $K_{\ell 4}$ decays
\cite{PT68,BCEG95}. They are: $s=(p_++p_-)^2$, invariant mass
squared of $\pi^+\pi^-$; $q^2=(k_++k_-)^2$, invariant mass squared
of $e^+ e^-$; $\theta_\pi$, the angle between the $\pi^+$
three-momentum and the $\eta$ three-momentum in the $\pi^+\pi^-$
rest frame; $\theta_e$, the angle between the $e^-$ three-momentum
and the $\eta$ three-momentum in the $e^+ e^-$ rest frame; $\phi$,
the angle between the $e^+ e^-$ and $\pi^+\pi^-$ planes in the
$\eta$ rest frame. Using these kinematic variables, a
CP-violating observable could be found by analyzing the $\phi$
distribution of the decay width
$\Gamma(\eta\rightarrow\pi^+\pi^-e^+e^-)$, which reads
\beq\label{CP1} {\cal A}_{\rm CP}=\langle sign(\sin\phi \cos
\phi)\rangle=\Frac{\int_0^{2\pi} \frac{d\Gamma(
\eta\rightarrow\pi^+\pi^-e^+e^-)}{d\phi}d\phi~ sign(\sin\phi \cos
\phi)}{\int_0^{2\pi}
\frac{d\Gamma(\eta\rightarrow\pi^+\pi^-e^+e^-)}{d\phi}d\phi}. \eeq
Now in terms of the invariant form factors defined in eq.
(\ref{Amp1}), and integrating over $\cos\theta_e$ and $\phi$, we
obtain \beqn\label{CP2} &&{\cal A}_{\rm CP}=\Frac{e^2}{3\cdot
2^{13}\pi^6
m_\eta^3\Gamma(\eta\rightarrow\pi^+\pi^- e^+ e^-)}\nonumber\\
&&\times \int d\cos\theta_\pi ~d s ~d q^2 ~\sin^2\theta_\pi
\Frac{s \beta^3 \lambda(s,m_\eta^2,q^2)}{q^2} {\rm
Re}\left[(E_+-E_-)M^*\right], \eeqn where \beq
\beta=\sqrt{1-\Frac{4m_\pi^2}{s}}, \eeq \beq
\lambda(a,b,c)=a^2+b^2+c^2-2 (a b+a c +b c). \eeq

The decay rate of $\eta\rightarrow\pi^+\pi^-\gamma$ has been
measured \cite{PDG} and is dominated by the magnetic transition
\cite{Amet01, Holstein01}. However, the prediction to the decay
rate from the lowest order chiral perturbation theory are somewhat
lower than the experimental value \cite{Amet01}.
Phenomenologically, one can adopt the following magnetic form
factor \cite{Picciotto}, constructed using chiral models, for the
magnetic transition of $\eta\rightarrow\pi^+\pi^-\gamma$
\beq\label{mag1} M=\Frac{e}{8\pi^2
f_\pi^2}\left(\Frac{1}{\sqrt{3}f_8}
\cos\theta-\sqrt{\Frac{2}{3}}\Frac{1}{f_0}\sin\theta\right)\times
\left(1-\Frac{3m_V^2}{m_V^2-s}\right), \eeq which yields the decay
width very close to the experimental value of
$\Gamma(\eta\rightarrow\pi^+\pi^-\gamma)$. Note that, in eq.
(\ref{mag1}), $m_V=770$ MeV is the vector-meson mass; $f_\pi=93$
MeV, $f_8=1.3 f_\pi$, and $f_0=1.1f_\pi$; $\theta=-20^o$ is the
angle of the $\eta-\eta^\prime$ mixing.

Since the CP asymmetry in eq. (\ref{CP2}) arises from the
interference between the $M$ and $E_\pm$ terms, in order to get
nonzero ${\cal A}_{\rm CP}$, we have to look for the possible
interaction which could yield the electric transitions of
$\eta\rightarrow\pi^+\pi^-\gamma^*\rightarrow \pi^+\pi^- e^+ e^-$.
As pointed out in Ref. \cite{GNW02}, it is easy to see that
bremsstrahlung electric amplitudes can be induced through the
$\pi^+\pi^-$ intermediate state which violates CP symmetry. Thus
it is straightforward to get \beq\label{IB1} E_+=\frac{2e
g_{\eta\pi^+\pi^-}}{q^2+2 q\cdot p_+}, \eeq and \beq\label{IB2}
E_-=-\frac{2e g_{\eta\pi^+\pi^-}}{q^2+2 q\cdot p_-}, \eeq where
$g_{\eta\pi^+\pi^-}$ is the effective coupling of
$\eta\rightarrow\pi^+\pi^-$ \cite{JS95,GNW02}. From the
experimental upper bound ${\rm
Br}(\eta\rightarrow\pi^+\pi^-)<3.3\times 10^{-4}$ \cite{PDG}, one
can get \beq\label{gvalue} |g_{\eta\pi^+\pi^-}|<0.12~{\rm MeV}.
\eeq Now using eqs. (\ref{CP2})---(\ref{IB2}) together with the
following scalar products of four-vectors \beqn\label{sp1} q\cdot
p_+=\frac{1}{4}(m_\eta^2-s-q^2)-\frac{1}{2}\beta
\lambda^{1/2}(s,m_\eta^2,q^2)\cos\theta_\pi,
\\
q\cdot p_-=\frac{1}{4}(m_\eta^2-s-q^2)+\frac{1}{2}\beta
\lambda^{1/2}(s,m_\eta^2,q^2) \cos\theta_\pi, \eeqn we can obtain
\beq\label{IBCP} |{\cal A}_{\rm CP}|=\Frac{3.7\times 10^{-4}}{{\rm
Br}(\eta\rightarrow\pi^+\pi^- e^+
e^-)}\cdot\Frac{|g_{\eta\pi^+\pi^-}|}{0.12~{\rm MeV}}\cdot
1.4\times 10^{-3}. \eeq The branching ratio of
$\eta\rightarrow\pi^+\pi^- e^+ e^-$ has been measured. Its value
listed in the present Particle Data Book \cite{PDG} is
\beq\label{PDG00} {\rm Br}(\eta\rightarrow\pi^+\pi^- e^+
e^-)=1.3^{+1.2}_{-0.8}\times 10^{-3}, \eeq which is from a very
old measurement \cite{RPC66}. The very recent measurement is from
CMD-2 Collaboration \cite{CMD2}: \beq\label{CMD-2} {\rm
Br}(\eta\rightarrow\pi^+\pi^- e^+
e^-)=3.7^{+2.5}_{-1.8}\pm0.3\times 10^{-4}. \eeq Obviously, the
situation for the observed ${\rm Br}(\eta\rightarrow\pi^+\pi^- e^+
e^-)$ is not very good due to the existing large uncertainty,
which of course needs to be further improved. As a
phenomenological analysis in the present paper, we use the central
value in eq. (\ref{CMD-2}) instead of the theoretical prediction
from eq. (\ref{Amp1}) to illustrate the numerical value for the
CP-violating asymmetry ${\cal A}_{\rm CP}$.

The experimental constraint on $g_{\eta\pi^+\pi^-}$ has been
obtained in eq. (\ref{gvalue}) by the limit of
Br($\eta\rightarrow\pi^+\pi^-$), which makes the upper bound of ${\cal
A}_{\rm CP}$ be about $10^{-3}$. Theoretically, in Ref.
\cite{JS95}, the CP-violating decay $\eta\rightarrow\pi^+\pi^-$
together with the effective coupling $g_{\eta\pi^+\pi^-}$ has been
studied both in the standard model based on CKM mechanism
\cite{CKM} and/or strong $\theta$ term lagrangian \cite{HCS} and
in the spontaneous CP violation model based on the Weinberg
mechanism \cite{Wein76}. Unfortunately,  the values of
$g_{\eta\pi^+\pi^-}$ given by the above mechanisms are much less
than the upper bound in eq. (\ref{gvalue}), which implies that CP
asymmetry in eq. (\ref{IBCP}) is very small and fully negligible.
As explained in Ref. \cite{GNW02}, this is not 
surprising at all because the CP-violating quantities such as $\epsilon$
and $\epsilon^\prime$ in the $K^0$ system, and the neutron
electric dipole moments $d_n$ have imposed very strong constraints
on the coupling of $\eta\rightarrow\pi^+\pi^-$. One of the good
choices to evade these constraints is that, we should search for
some unconventional sources of CP violation which do not
contribute directly to $\epsilon$, $\epsilon^\prime$, and $d_n$
and yet has a contribution to the decay
$\eta\rightarrow\pi^+\pi^-\gamma(\gamma^*)$. A type of such
operators, flavor-conserving CP violating four-quark operators
involving two strange quarks together with combinations of other
light quarks, have been proposed in \cite{GNW02}, which can be
explicitly written as follows \beq\label{operator} {\cal
O}=\Frac{1}{m_\eta^3}G\bar{s}i\sigma_{\mu\nu} \gamma_5 (p-q)^\nu
s\bar{\psi}\gamma^\mu \psi, \eeq where $\psi$ denotes up or down
quarks, $G$ is a dimensionless parameter originating from
short-distance physics and it can be taken real due to the CPT
invariance. In order to evaluate the contribution from eq.
(\ref{operator}) to $\eta\rightarrow\pi^+\pi^-\gamma$, the authors
of Ref. \cite{GNW02} used a factorization approximation that the
$\pi^+\pi^-$ part is from $\bar{\psi}\gamma^\mu \psi$ and the
$\eta\gamma$ transition involves only part containing strangeness.
Following the same procedure, it is easy to get the electric form
factors of $\eta\rightarrow\pi^+\pi^- e^+ e^- $ as \beq\label{Ep2}
E_+\sim \Frac{e F(s) G}{m_\eta^3}(q^2+2 q\cdot p_-), \eeq and
\beq\label{Em2} E_-\sim-\Frac{e F(s) G}{m_\eta^3}(q^2+2 q\cdot
p_+), \eeq where $F(s)$ parameterizes the form factor of the
$\eta\rightarrow\gamma$ transition. It has been estimated as
$F(s)\sim F(0)\sim 0.19$ by the authors of Ref. \cite{GNW02} using
the light front quark model \cite{GLZ}. From eqs. (\ref{CP2}),
(\ref{mag1}), (\ref{Ep2}), and (\ref{Em2}), we obtain
\beq\label{NewCP} |{\cal A}_{\rm CP}|\sim \Frac{3.7\times
10^{-4}}{{\rm Br}(\eta\rightarrow\pi^+\pi^- e^+ e^-)}\cdot 2.0
\times 10^{-2}~ |G|.
 \eeq
If $G$ is $O(1)$, ${\cal A}_{\rm CP}$ could be about $2.0 \times
10^{-2}$, which is one order larger than the upper bound in eq.
(\ref{IBCP}). Note that, for $G\sim$ 1, we still have $|M|$ is
much large than $|E^\pm|$, which does not essentially change the
prediction from eq. (\ref{mag1}) for the decay rate of
$\eta\rightarrow\pi^+\pi^-\gamma$. On the other hand, as mentioned
in Ref. \cite{GNW02}, since the operator in eq. (\ref{operator})
cannot directly generate the decay $\eta\rightarrow\pi^+\pi^-$ and
also cannot induce $d_n$ either, the CP-violating contributions
from the operator in eq. (\ref{operator}) are free of the strong
constraints from $\epsilon$, $\epsilon^\prime$, and $d_n$. Hence,
there are basically no direct constraints for $G$ from both low
and high energy experiments. Thus in principle, $G\sim O(1)$ is
allowed, and a sizable CP asymmetry ${\cal A}_{\rm CP}$ could be expected.

In this paper, we are concerned about the general aspects of the
CP-violating asymmetry in $\eta\to\pi^+\pi^- e^+ e^-$, and
our study seems to indicate that there exists the possiblity of observing
such effects. One might suspect
that $G\sim O (1)$ is over-estimated. This is not very surprising because
generally  the CP asymmetry of this decay
is strongly suppressed in many conventional mechanisms. 
We should further explore how to realize the uncoventional
operators of eq. (\ref{operator}) in some explicit theoretical models, in
which $G$ could be enhanced up to $O(1)$ or not much smaller than $1$.
This is not the task of the present paper, and it will be considered in
the future study. On the other hand,
the measurement of the CP asymmetry ${\cal A}_{\rm CP}$  in $\eta\to
\pi^+\pi^- e^+ e^-$ would  provide an interesting limit on $G$, thus
further impose the constraints on the theoretical models.    

The CP-violating effects in $\eta\rightarrow\pi^+\pi^-\gamma$,
which is generated from the interference between the magnetic and
electric amplitudes of the decay with explicit photon
polarization, has been studied in Ref. \cite{GNW02}, and a sizable
CP-violating effect could be induced when the contributions from
an unconventional CP-violating interaction in terms of a tensor
type operator in eq. (\ref{operator}) are considered. Note that
such CP-violating effects will be invisible as long as the
polarization of the photon is not observed. As an alternative to
measuring the photon polarization, we consider in the present
paper the decay $\eta\rightarrow\pi^+\pi^- e^+ e^-$ resulting from
the internal conversion of the photon into an $e^+e^-$ pair, and the
CP-violating effects hidden in the polarization
of the photon now can be translated into the CP asymmetry ${\cal
A}_{\rm CP}$ in the angular correlation of the $e^+ e^-$ plane
relative to the $\pi^+\pi^-$ plane.  The upper bound of ${\cal A}_{\rm
CP}$ is about $10^{-3}$ by using the experimental limit of
$Br(\eta\to
\pi^+\pi^-)$; more interestingly, the value of ${\cal A}_{\rm CP}$ might
be enhanced up to $O(1\%)$ in some unconventional scenarios. Hopefully,
for the future high statistics experiments in the $\eta$ factories, this
asymmetry could be a useful CP-violating observable both to increase our
knowledge on CP violation and to search for new physics effects
beyond the standard model.

\vspace{0.4cm}
\section*{Acknowledgments}

\indent 
The author wishes to thank Chao-Qiang Geng and Paul Singer for very
helpful communications. This work is supported in part by the NSF of China
under Grant No. 19905008.

\end{document}